\documentclass[runningheads]{llncs}

\usepackage{pdfpages}
\usepackage{circledsteps}
\usepackage{multirow}
\usepackage{tabularx}
\usepackage{float}
\usepackage{subfig}
\usepackage{amsfonts}
\usepackage{mathtools}

\usepackage{cryptocode}
\usepackage[pdf]{graphviz}
\usepackage{forest}
\usepackage{xcolor}
\usepackage{tikz}
\usepackage[group-separator={,}]{siunitx}
\usetikzlibrary{arrows.meta, chains, positioning, shapes.multipart}
\tikzset{
    hashtable/.style = {rectangle split, rectangle split parts=6, draw, inner sep=8pt,minimum height=2em},
    listnode/.style = {rectangle split, rectangle split horizontal, rectangle split parts=2, draw, on chain}
}

\makeatletter
\DeclareRobustCommand{\rvdots}{%
  \vbox{
    \baselineskip4\p@\lineskiplimit\z@
    \kern-\p@
    \hbox{.}\hbox{.}\hbox{.}
  }}
\makeatother

\usepackage{xspace}
\newcommand{\prot}{\textsc{Accept}\xspace}

\begin{document}

\title{A Limitlessly Scalable Transaction System}

\author{Max Mathys \and Roland Schmid \and Jakub Sliwinski \and Roger Wattenhofer}

\institute{ETH Zürich}

\maketitle

\begin{abstract}
We present \prot, a simple, asynchronous transaction system that achieves perfect horizontal scaling.

Usual blockchain-based transaction systems come with a fundamental throughput limitation as they require that all (potentially unrelated) transactions must be totally ordered. Such solutions thus require serious compromises or are outright unsuitable for large-scale applications, such as global retail payments.

\prot provides efficient horizontal scaling without any limitation. To that end, \prot satisfies a relaxed form of consensus and does not establish an ordering of unrelated transactions. Furthermore, \prot achieves instant finality and does not depend on a source of randomness.
\end{abstract}

\section{Introduction}


The financial world is changing around the globe. With the rise of digital cryptocurrencies like Bitcoin~\cite{nakamoto2008bitcoin} and Ethereum~\cite{wood2014ethereum}, the pressure on the traditional banking system to implement a digital currency on its own is rising. Due to the well-known limitations of permissionless blockchain systems thus far, such a digital currency is mostly envisioned based on permissioned, byzantine fault-tolerant ledger technology. Furthermore, as both central and commercial banks have no interest in a fully distributed solution which can hardly be regulated, the employment of a permissioned system constitutes a good fit to establish such a digital currency.

Previously, revolutionizing the global financial infrastructure was envisioned on the basis of byzantine agreement protocols. Such systems are popular because they provide a reliable and robust way of transferring funds between participants by establishing a total order of all transactions. However, ordering all transactions has proven to be a throughput-limiting factor for these systems, which only achieve throughputs up to tens of thousands of transactions per second even in lab environments~\cite{stathakopoulou2019mir}. Despite byzantine agreement systems being optimized for high throughput, they only recently matched the demands of leading credit card providers; hence, this technology does not seem future-proof in our increasingly digital world.

\paragraph{Our Contribution:}
We present a system based on A Cheaper Consensus for Efficient, Parallelizable Transactions (\prot), that features:
\begin{itemize}
    \item \textbf{Limitless Scalability:} \prot does not order transactions that do not depend on each other. Thus, any number of parallel transactions can be processed given sufficient amount of hardware.
    \item \textbf{Instant Finality:} \prot confirms transactions in 1 round-trip time to the validator nodes. Confirmed transactions are final and cannot be reversed.
    \item \textbf{Asynchrony:} \prot does not require any network timing assumptions. An adversary having complete control over the network can halt the progress of the system (by simply disabling communication), but otherwise cannot trick the participants in any way, such as reversing a confirmed transaction or pretending that an impermissible transaction is confirmed.
    \item \textbf{Independent Validators:} Validator nodes only need to provide a basic API to accept new transactions, verifying them internally and returning a valid signature in case of success. This makes the nodes' implementations independent, as they only need to interpret other nodes' (and clients') signatures. Validators can parallelize and balance workload as they individually see fit, without changing the way they interact with the system.
    \item \textbf{Simplicity:} \prot does not rely on randomness, is easy to comprehend and to implement. Hence, given some deployment scenario, the validators can easily implement the protocol themselves, thus achieving fault-tolerance with respect to software bugs in the code. In contrast, existing (complicated) permissioned blockchain systems rely on reference implementations that are used as a black-box, where a single bug can compromise the entire system.
\end{itemize}
We demonstrate \prot's horizontal scaling on common server hardware and showcase its transaction throughput to be orders of magnitudes above byzantine agreement systems.

\section{Model and Preliminaries}

\prot is maintained by $n$ different agents called \textit{validators}. Similarly to other byzantine fault tolerant systems, we assume up to $f = \frac{n-1}{3}$ of the validators are adversarial and behave in an arbitrary way. Any set of $2f+1$ validators is called a \textit{quorum}. In addition, an arbitrary number of \textit{clients} interact with the system by issuing and receiving transactions.

The network is asynchronous: The adversary controls the network, dictating when messages are delivered and in what order. There is no bound on the time it might take to deliver a message. Under such weak network requirements, an adversary delaying the delivery of messages can delay the progress of the protocol, but otherwise will not be able to interfere.

We assume the functionality of digital signatures where a public key allows to verify a signature of the associated secret key. We also assume cryptographic hashing, where for every message a succinct, unique hash can be computed. Apart from these primitives, \prot is completely deterministic.

\paragraph{Security \& Threat Model.}
Validators and clients hold public/private key pairs. All participants know the public keys of all validators.

The adversary knows the protocol and controls all adversarial validators, and any number of clients. The adversary controls the network and can delay, replay, reorder messages etc. The adversary does not know the private key of any correct participant.

\section{Protocol}

The \prot protocol differentiates two main roles:
\begin{itemize}
    \item \textbf{Validators:} Validators are agents that \textit{verify} and \textit{sign} transactions.
        There is a fixed number of validators. Validators do not have to exchange messages with each other. Validators can be sharded across multiple servers to increase the throughput of the system.
    \item \textbf{Clients:} Clients are end users of the system who \textit{issue}/\textit{receive} transactions.
        The system supports an arbitrary number of clients in the system. Clients possess funds that can be sent to other clients via a transaction. Clients may follow the protocol correctly or not; however, it is only guaranteed that they can spend their funds if following the protocol.
\end{itemize}
Transactions are processed by the system in the UTXO model~\cite{delgado2018utxo}. The initial state of the system, called {\it genesis}, consists of (number, public key) pairs, where the number represents the available funds, and the public key identifies the party able to spend them. These pairs are called unspent transaction outputs (UTXO).

If a client (sender) wishes to transfer funds to another client (recipient), the sender issues a transaction. The transaction contains input UTXOs that the sender can spend. The sender specifies output UTXOs, where inputs and outputs sum up to the same amount of funds. For example, a single transaction can specify two outputs, where one output represents the transacted amount and includes the public key of the recepient, and the second output represents the change and includes the public key of the sender. Ultimately, the sender signs the transaction with the private key(s) corresponding to the inputs.

\paragraph{Transaction Pipeline.}
\prot performs four steps to confirm a transaction from client~$c_1$ to client~$c_2$:
\begin{enumerate}
    \item \textbf{Issuing a Transaction:} The client~$c_1$ composes and signs a transaction. Client~$c_1$ sends the transaction to the validators. If the inputs of the transaction are not part of genesis, the client also sends the confirmations of the inputs to the validators.
    \item \textbf{Verifying a Transaction:} Each validator~$v_i$ verifies the transaction signature and inputs' confirmations (if not part of genesis). Also, $v_i$ checks that it has not validated any transaction spending the same inputs thus far.
    \item \textbf{Signing a Transaction:} If the transaction is valid, validator~$v_i$ signs the transaction and returns the signature to $c_1$.
    \item \textbf{Finalizing a Transaction:} A set of $2f + 1$ signatures of distinct validators constitutes a \textit{confirmation} of the outputs. The client $c_1$ can show the corresponding transaction output and the confirmation to $c_2$ to prove the transfer took place. The client~$c_2$ accepts the transfer after verifying the output and the confirmation.
\end{enumerate}

\subsection{Correctness}
\paragraph{Double-spending.}
Suppose some execution of the protocol produced two confirmed transactions $t_1$ and $t_2$ that spend the same output. Each confirmed transaction is signed by a validator quorum. Since the adversary controls at most $f$ validators, at least $f+1$ correct validators signed $t_1$ and $t_2$. Since there are $2f+1$ correct validators, some correct validator $v$ signed both $t_1$ and $t_2$. However, when signing a transaction, correct validators check whether they have not signed any of the inputs previously -- a contradiction.

\paragraph{Finality.}
Given a confirmed UTXO, it can only be invalidated if some validators observe a transaction that spends the UTXO. Only the owner of the UTXO can sign such a transaction.

\paragraph{Liveness.}
Any $2f+1$ validators can confirm any transaction. Since at least $2f+1$ validators are correct, the correct validators can confirm transactions if the adversary refrains from participating.

\subsection{Signature Protocol}\label{sec:sigprotocol}
Naively, validators can sign each transaction separately and verify separate signatures for each output. In addition to this naive approach, we design two different protocols for batch-processing the transactions, thereby vastly improving the system's performance: the Merkle scheme and the BLS scheme.

\subsection{Merkle Scheme}
\label{sec:merkle}
In the Merkle scheme, validators combine many transactions into Merkle trees and only sign the root, effectively signing many transactions at once.

\paragraph{Signing.} By \textit{pooling} multiple signing requests, the validator collects a large number $p$ of unsigned outputs. The hashes of the outputs $h_0(i) = h(o_i)$ are hashed in pairs $h_1(i) = h(h_0(i), h_0(i+1))$, the resulting hashes are hashed in pairs $h_2(i) = h(h_1(i), h_1(i+2))$ and so on, to create a complete binary tree of hashes, where the leaves are the hashes of the UTXOs to be signed.

The validator signs the root of the tree. For each output $o_i$, the validator will return to the issuer of $o_i$ the hashes needed to compute the path from $h(o_i)$ to the root: the hash $x_1$ to compute $h_1(i) = h(h_0(i), x_1)$, the hash $x_2$ to compute $h_2(i) = h(h_1(i), x_2)$, and so on. These hashes $x_j$ together with the signature of the tree root constitute $o_i$'s signature in the Merkle scheme (see Appendix~\ref{sec:merkle-appendix}).

When a validator signs $p$ outputs, only one signing operation is executed (compared to $p$ signing operations with the naive scheme). However, many hash operations must be performed for both signing and verification in this scheme.

\paragraph{Signature verification.}
As in the naive scheme, for an output to be confirmed, there must be $>\frac{2}{3}n$ signatures from different validators. We verify each Merkle signature $s_i=(\{x_j\}, s_i^{\textit{root}})$: we reconstruct the hash path to the root using the $x_j$'s and verify the signature of the root. There are $p$ signatures with the same Merkle root; hence, the verification result can be cached in memory by the validator such that the validator only verifies the root signature the first time it is observed. For all $p-1$ subsequent encounters of the root, it suffices that the validator performs $\log(p)$ hash operations.

\paragraph{Optimal Merkle Tree Size.}
If the Merkle tree is very large, the hashing time dominates the verification and signing process. However, if the Merkle tree size is small, the cryptographic operations dominate. We omit the work not related to cryptographic operations or hashing, and estimate the optimal tree size.

Let $q$ be the quorum size of the system, $N$ the number of leaves in the Merkle tree, $c_h$, $c_s$, $c_v$ the costs of hashing, signing and verification. Let $C_{naive}$ and $C_{merkle}$ denote the average time cost incurred by a validator to process one UTXO under the naive and Merke signature protocols. Each validator signs a UTXO once, and later has to verify the signatures constituting that UTXO's confirmation. Hence, the expected cost for the naive scheme is
$C_{naive} = q \cdot c_v + c_s$.

The expected cost of the Merkle scheme is
$$
C_{merkle} =
    \underbrace{
        q \left(c_h \log N + \frac{c_v}{N}\right)
    }_\text{Verification}
    +
    \underbrace{
       \frac{2N c_h + c_s}{N}
    }_\text{Signing}
    = \log N (q c_h) + \frac{1}{N} (q c_v + c_s) + 2 c_h.
$$
The expression is minimized by $N = \frac{q c_v + c_s}{q c_h} \ln 2$. For example, if the relative operation costs are around $c_h = 1$, $c_s=63$, $c_v=107$ and there are 10 validators (quorum size 7), we estimate the optimal number of leaves in the Merkle tree at $N^{\textit{optimal}} \approx 80.4$.
For binary trees ($\log N$ being integer), either a Merkle tree with 64 or 128 leaves is optimal.

\subsection{BLS Scheme}
\label{sec:bls}

BLS \cite{boneh2004short} is a signature scheme where signatures can be aggregated. One verificaton operation on an aggregated signature can be used to verify all constituent signatures at once.
Combining the properties of BLS with Shamir's secret sharing, it is possible to construct a threshold signature scheme. In this scheme, each validator possesses a different private key and signs transactions for clients individually. Once a client obtains at least $2f+1$ signatures for their transaction, they can use the signatures to compute a unique {\it master} signature of their transaction. The master signature is unique and the same, irrespective of which $2f+1$ validators' signatures were used to compute it.

Due to space constraints, we do not describe the working of this threshold scheme in detail (the reader can find an instructive description at \cite{blsdashblog}).

The BLS sigantures are relatively costly to produce and verify. However, the scheme comes with the great advantage that aggregated BLS threshold signatures have a constant verification time, irrespective of the number of validators in the system.

Each validator receives a BLS private key, and the corresponding public keys are publicly known. The master public key is publicly known (or can be computed given the validator public keys). The validators sign the outputs of the clients with their BLS keys, similarly to the naive scheme. The validators' signatures function on their own as usual, so clients can verify that they receive the correct signatures. After receiving $2f+1$ signatures, a client can compute the unique master signature for their outputs. Most importantly, when validators receive transactions to be signed, they only need to verify one master signature for each input of the transaction.

\section{Implementation}
The validator node and client are written in Go due to its performance and ease of parallelization.
The implementation features the three different signature protocols described in Sections \ref{sec:sigprotocol}-\ref{sec:bls}: the naive scheme, the Merkle scheme and the BLS scheme. For signing and verification, the naive and Merkle schemes use EdDSA with Curve25519 (Ed25519). A Go library \cite{golib1} provides bindings to ed25519-donna \cite{ed25519donna}. Ed25519-donna is written in C++ and provides a fast implementation of the Ed25519 public-key signature system \cite{ed25519}. Batch verification can be used for greater throughput. The BLS scheme is implemented using the \texttt{herumi/bls} \cite{herumi} with Go-bindings \cite{herumibindings}.

\paragraph{Benchmarking Merkle tree size.} The signing and verification times for the Merkle signature scheme have been measured on the AWS reference instance for different Merkle tree sizes and 10 validators. The observed global minimum around 64-128 leaves matched the conclusion from Section \ref{sec:merkle}.

\subsection{Cryptographic Scheme Comparison}
The benchmark of the cryptographic operations executed on a single core on the reference instance is denoted in Table \ref{tab:crypto_bench}.

\begin{table}
    \centering
    \def\arraystretch{1.2}

    \begin{tabularx}{0.7\textwidth}{>{\hsize=0.5\hsize\linewidth=\hsize\arraybackslash}X|>{\hsize=1.5\hsize\linewidth=\hsize\centering\arraybackslash}X|>{\hsize=1\hsize\linewidth=\hsize\centering\arraybackslash}X}
        \textbf{Scheme} & \textbf{Operation} & \textbf{ns per signature} \\
        \hline
        \multirow{3}{*}{Naive} & Signing & \num{29967} \\
         & Verifying, single & \num{100663} \\
         & Verifying, batch of 64 & \num{51247}  \\
         \hline
        \multirow{3}{*}{Merkle} & Signing & \num{2709} \\
        & Verifying, no caching & \num{106771} \\
        & Verifying, cached & \num{6473} \\
        \hline
        \multirow{2}{*}{BLS} & Signing & \num{640205} \\
         & Verifying & \num{1918578} \\
    \end{tabularx}\\[2mm]
    \caption{Benchmarks for relevant cryptographic operations for each scheme. Fort the Merkle scheme, the number of leaves is 64.}
    \label{tab:crypto_bench}
\end{table}
If multiple signatures are verified in one batch, EdDSA can take advantage of x86 SIMD instructions. This gives verification a speedup of up to $\approx 2$.
Assuming that the Merkle tree has 64 leaves, the Merkle scheme is about a magnitude faster than the naive scheme.
Since verification of threshold BLS signatures is constant for any number of validators, we find BLS to be faster than the naive and Merkle schemes
if the number of validators is $> 37$ and $> 475$, respectively. Moreover, for larger quorum sizes, the naive and Merkle schemes yield larger transaction confirmations, whereas BLS' confirmations have a constant size.

\subsection{Storage}
\label{sec:storage}

Each validator keeps track of outputs they signed as spent.
The spent outputs are stored in a thread-safe and efficient hash map. Golang’s built-in thread-safe hash map, sync/map, exhibits excessive lock usage and coroutine blocking; hence, we implemented a purpose-built hash map (see Appendix~\ref{sec:hashmap-appendix}). Running on the AWS reference instance with one coroutine per processor, Golang's implementation reaches $7.04 \cdot 10^5$ inserts per second, and our implementation about $3.30 \cdot 10^7$ inserts per second.

\subsection{Sharding}
\label{sec:sharding}

Crucially, validators can easily shard their workload among multiple machines. Clients are assigned to different shards based on their public key; in this implementation, inputs corresponding to different public keys cannot be mixed in one transaction. In our implementation, the assignment of clients to shards is publicly known and clients request the machines they are assigned to. Alternatively, validators could use some load balancing approach.

\section{Evaluation}


We performed end to end benchmarks of the system using the Merkle scheme with servers rented at AWS.
We tested with 4, 10, and 28 validators. For each number of validators, we performed an experiment with 1, 2, and 4 shards (machines) for each validator. The server instances used in these benchmarks were c3.8xlarge with 32 virtual CPU threads and 60 GiB of RAM. Each test consisted of a preset number of client servers (corresponding to the expected throughput) generating transactions and sending them to appropriate shards of the validators. The duration of each test was ten minutes. The results of the experiment are presented in Table \ref{tab:results}. The average CPU utilization observed was $80.9\%$.

\begin{figure}
    \centering
    \includegraphics[width=0.7\textwidth]{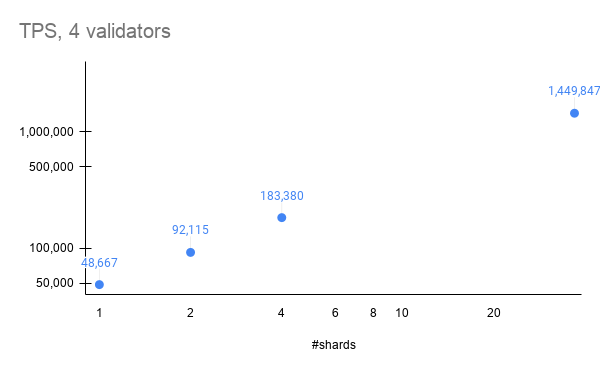}
    \caption{TPS scalability results for 4 validators, log scales.}
    \label{fig:4vals}
\end{figure}

We performed an additional experiment with 4 validators and 37 shards per validator (and otherwise the same setup), yielding a throughput of 1,449,847 transactions per second on average, with an average CPU utilization of $71.9\%$. The results of this experiment are presented with logarithmic scales in Figure~\ref{fig:4vals}.

\begin{table}
    \centering
    \def\arraystretch{1.2}
    \begin{tabularx}{0.7\textwidth}{>{\centering\arraybackslash}X|>{\centering\arraybackslash}X|>{\centering\arraybackslash}X|>{\centering\arraybackslash}X}
        \#validators & 4 & 10 & 28 \\
        \hline
        1 shard & \num{48667} & \num{34451} & \num{22015} \\
        2 shards & \num{92115} & \num{72079} & \num{44764} \\
        4 shards & \num{183380} & \num{141767} & \num{80917} \\
    \end{tabularx}\\[2mm]
    \caption{Average transactions per second in the experiment.}
    \label{tab:results}
    \vspace*{-0.8cm}
\end{table}

\section{Related Work}

The first work to suggest a simpler, consensus-free approach to processing transactions in a permissioned system was Gupta~\cite{gupta}. Gupta focuses on providing verifiable audit trails. We employ a similar transaction confirmation principle and focus on designing and implementing a system around horizontal scaling without loss of efficiency.

ABC~\cite{abc} describes how a consensus-free system can be managed similarly to proof-of-stake blockchains, thus extending the approach to the permissionless setting. ABC also contributes some features of the protocol, such as pruning redundant contents from the blockchain, and discusses some economic aspects of the permissionless setting. However, ABC does not provide an implementation.

FastPay~\cite{fastpay} is concurrent work that provides an implementation of a permissioned settlement system and focuses on interfacing with a preceding, primary system. FastPay is similar in spirit to \prot, but more complicated in crucial aspects; for example, FastPay employs a two-phase confirmation protocol that complicates the interaction between the validators. However, FastPay does not implement efficiency improvements similar to those described in Sections \ref{sec:merkle} and \ref{sec:bls}. Parallel processing in FastPay's implementation is only process-based, and hence does not exemplify multi-machine sharding. The performance is also not reported clearly: benchmarks are presented for the two confirmation phases separately, whereas one exhibits a bottleneck in the process-based implementation.

\bibliographystyle{splncs04}
\bibliography{accept}

\newpage
\appendix

\section{Concepts}

This section will clarify the concepts of Merkle signatures and hash maps. Merkle signatures are used in Section~\ref{sec:merkle} to improve the efficiency of validators. A custom hash map is developed to improve the performance of the UTXO store, as described in Section~\ref{sec:storage}. 

\subsection{Merkle Signatures}
\label{sec:merkle-appendix}

A signature scheme based on Merkle trees is used to optimize the performance of validators when creating and verifying signatures.

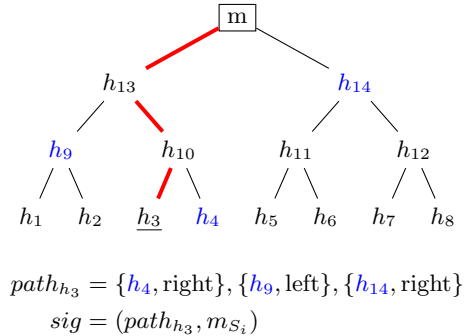
\begin{figure}[H]
    \centering
    \begin{forest}
    [m,draw,for tree={}
        [$h_{13}$,edge={red,line width=1.5pt}
            [$h_9$,blue
                [$h_1$]
                [$h_2$]
            ]
            [$h_{10}$,edge={red,line width=1.5pt}
                [\underline{$h_3$},edge={red,line width=1.5pt}]
                [$h_4$, blue]
            ]
        ]
        [$h_{14}$,blue
            [$h_{11}$
                [$h_5$]
                [$h_6$]
            ]
            [$h_{12}$
                [$h_7$]
                [$h_8$]
            ]
        ]
    ]
    \end{forest}
    \begin{align*}
        path_{h_3} &= \{\textcolor{blue}{h_4}, \text{right}\}, \{\textcolor{blue}{h_9}, \text{left}\}, \{\textcolor{blue}{h_{14}}, \text{right}\}\\
        sig &= (path_{h_3}, m_{S_i})
    \end{align*}
    \caption{Merkle tree and Merkle signature for hash $h_3$.}
    \label{fig:merkle}
\end{figure}

\paragraph{Signing.} A validator $S_i$ collects $n$ hashes to sign where $n=2^k, k \in \mathbb{N}$. The hashes are combined into a Merkle tree. The validator $S_i$ signs the Merkle root $m$ using EdDSA, denoted as $m_{S_i}$. For each hash $h_i, i \in \{1,..,n\}$, the validator calculates the Merkle path and outputs $path_{h_i}$. $path_{h_i}$ consists of the hashes and side (left or right) of the nodes from $h_i$ to the Merkle root (in blue). The resulting signature for $h_i$ by $S_i$ is the tuple $(path_{h_i}, m_{S_i})$. An example can be seen in Figure~ \ref{fig:merkle}.

Signing $n$ hashes using Merkle signatures is more efficient than signing $n$ hashes separately: the cost of an EdDSA signing operations $c_s$ is much higher than the cost of a hash operation $c_h$. The cost of signing $n$ hashes with Merkle signatures is $2n\cdot c_h + c_v$ whereas the cost of signing $n$ hashes separately (without Merkle signatures) is $n \cdot c_v$.

\paragraph{Verifying.} First, the Merkle root $m'$ is reconstructed from the path $path_{h_i}$. If $path_{h_i}$ is a valid path, $m'$ matches $m$. Finally, $m_{S_i}$ is verified using EdDSA.

Verifying $n$ hashes using Merkle signatures is also more efficient than verifying $n$ hashes separately because the cost of an EdDSA verification operation is high, even higher than an EdDSA signing operation (and thus also higher than the cost of a hashing operation). The Merkle root can be reconstructed with a cost of $\log_2(n) \cdot c_h$ where $c_h$ is the cost of hashing. The Merkle root $m$ only has to be verified for the first encountered Merkle signature, after that, the result of the verification can be cached, making this signature perform very well.

\subsection{Hash Maps}
\label{sec:hashmap-appendix}

Hash maps are used to efficiently calculate set membership in the context of spent UTXO identifiers. Elements $e \in E$ (in our case $id_i$) are hashed into the hash range $H = \{0,...,l\}$ and used as an index to an array of linked lists. A linked list at index $i \in \{0, ..., l\}$ contains all items $e$ where $\mathrm{hash}(e) = i$. The linked list is then traversed.

Hashing and indexing are implemented as a lock-free operation; traversing and modifying a linked list are protected by a mutex. If $l$ is large enough, the probability of hash collisions is small, minimizing lock contention and the length of the linked list.

\begin{figure}
    \centering
    \begin{tikzpicture}[node distance = 0mm and 9mm, start chain = A going right] 
    \node (id1) {$\text{hash}(id_1)$};
    \node[below=0pt of id1] (id2) {$\text{hash}(id_2)$} ;
    \node[below=0pt of id2] (id3) {$\vdots$};
    \node[hashtable, right=2cm of id2] (H) {0 \nodepart{two} 1 \nodepart{three} $\rvdots$ \nodepart{four} k \nodepart{five} $\rvdots$ \nodepart{six} $l$ };
    \node[listnode, right=of H.four east] {$id_1$: true};
    \node[on chain] {$\hdots$};
    \draw[Circle-stealth] (H.four east) edge (A-1)
        (A-1.two |- A-1.two east) edge (A-2)
        (id1.east |- id1.east) edge (H.four west);
    \end{tikzpicture}
    \caption{A hash map, where $id_1$ indexes into $\text{hash}(id_1)=k$}
    \label{fig:hashmap}
\end{figure}
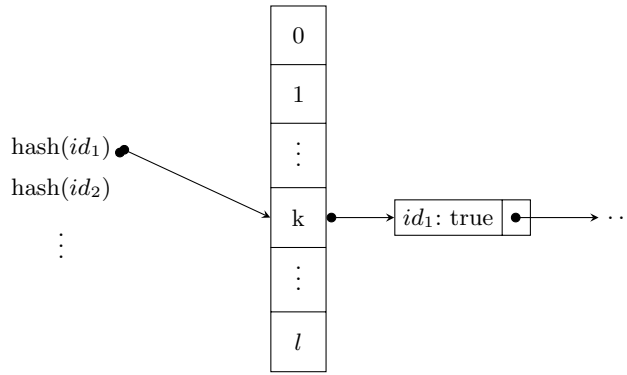

\end{document}